\documentclass[prb,
twocolumn,
superscriptaddress,showpacs,amsmath,amssymb,longbibliography]{revtex4-1}

\usepackage{xcolor}
\usepackage{graphicx}
\usepackage{hyperref}

\begin{document}

\title{Quantum turbulence and Planckian dissipation}

\author{G.E.~Volovik}
\affiliation{Low Temperature Laboratory, Aalto University,  P.O. Box 15100, FI-00076 Aalto, Finland}
\affiliation{Landau Institute for Theoretical Physics, acad. Semyonov av., 1a, 142432,
Chernogolovka, Russia}

\date{\today}

\begin{abstract}
The notion of the Planckian dissipation is extended to the system of the  Caroli-de Gennes-Matricon discrete energy levels in the vortex core of superconductors and fermionic superfluids. In this extension, the Planck dissipation takes place when the relaxation time 
$\tau$ is comparable with the quantum Heisenberg time $t_H=\hbar/\Delta E$, where $\Delta E$ is  the interlevel distance in the vortex core (the minigap). This type of Planck dissipation has two important physical consequences.  First, it determines the regime, when the effect of the axial anomaly becomes important. The anomalous spectral flow of the energy levels along the chiral branch of the Caroli-de Gennes-Matricon states becomes important in the super-Planckian region, i.e. when $\tau <\hbar/\Delta E$. Second, the Planck dissipation separates the laminar flow of the superfluid liquid at $\tau<\hbar/\Delta E$ and the vortex turbulence regime at $\tau>\hbar/\Delta E$.
\end{abstract}
\pacs{
}

\maketitle 
 
\section{Introduction}

For quantum system in equilibrium at temperature $T$ one can introduce the corresponding time unit, $t_T=\hbar/T$. In some condensed matter systems $t_T$ has physical significance, which suggested the introduction of the term  "Planckian dissipation", see e.g. Refs.\cite{Zaanen2021,Zaanen2019,Taillefer2019,Taillefer2021,Taillefer2022} and criticism in Ref. \cite{Sadovskii2021}. The dissipation is called Planckian, when the relaxation time  $\tau$ is comparable with  $\hbar/T$.

 In principle, the other  types of  dissipation can be also called Planckian, when the relaxation time $\tau$ is comparable with 
the other quantum time scales. Then, in general, the Planckian dissipation takes place, when the time scales related to dissipation (Thouless time\cite{Thouless1974} $t_{\rm H}$, elastic scattering time $\tau_{\rm el}$,  inelastic scattering time $\tau_{\rm inel}$, spin-relaxation time, etc.) are compared with the characteristic energy scale (temperature $T$, kinetic energy per particle, gap in the energy spectrum, interlevel distance $\Delta E$ in mesoscopic systems and in the vortex core, Zeeman splitting of the energy levels, etc.).
 For example:
 
(i) Dissipation can be called Planckian, when $\hbar/\tau$ is comparable with the interlevel distance, $\hbar/\tau \sim \Delta E$.
In quantum systems with discrete energy spectrum the corresponding time scale, $t_{\rm H}=\hbar/\Delta E$, is called the Heisenberg time. The corresponding Planckian dissipation, when $\tau \sim t_{\rm H}$,  has an important physical significance in the vortex dynamics  in superconductors and fermionic superfluids.  For the discrete energy levels in the core of vortices\cite{Caroli1964} the ratio $Re_v=\tau /t_{\rm H}$ serves as the special Reynolds number, which determines the onset of the turbulence of quantized vortices.\cite{Finne2003,Finne2006}  In this paper we shall discuss 
the relation between the quantum turbulence and the corresponding Planckian dissipation.

Let us also mention the other "Planckian" issues:

(ii) Dissipation can be called Planckian, when the Heisenberg time $t_{\rm H}$ is comparable with the Thouless time $t_{\rm Th}$.
Thouless time\cite{Thouless1974} is the time scale for a particle to diffuse through a disordered mesoscopic sample across a box of size $L$ and reach the boundaries: $t_{\rm Th} = L^2/D$, where $D$ is diffusion.

 (iii) Dissipation can be called Planckian, if the kinematic viscosity $\nu$ is comparable with the circulation quantum $2\pi \hbar/m$ of the superfluid velocity, where $m$ is the mass of the corresponding boson (mass of $^4$He atom in superfluid $^4$He, mass of two $^3$He atoms  in superfluid $^3$He).\cite{Sreenivasan2014} This is related to the notion of the minimal viscosity,\cite{Policastro2001,Kovtun 2005} and also to turbulence.\cite{Zaanen2019}

In what follows, we shall keep the Planck constant $\hbar$. In principle, one can use $\hbar=1$, if $\hbar$ is the fundamental quantity. However, in this case it is difficult to resolve between the quantum levels and the classical frequencies.
One can also absorb $\hbar$ into the space-time metric.\cite{Volovik2009} In this case the quantum and classical quantities are well resolved.

\section{Vortex core fermions, vortex turbulence and Planck dissipation}
\label{CoreFermions}

Here we consider vortices moving in the fermionic systems, such as Abrikosov vortex in the $s$-wave superconductors and vortices in superfluid phases of $^3$He.\cite{VW}
The vortex is the autonomous subsystem. It moves in the superfluid/superconducting environment, which is the bath of the Bogoliubov quasiparticles. The bath has the temperature $T$, which enters as the parameter for the vortex core subsystem. In principle, the temperature $T$ of the core fermions can be different from the bulk temperature,\cite{Silaev2013} but at the moment we do not discuss such situation here.

The original Planck dissipation\cite{Zaanen2021,Zaanen2019}  may take place only in case when $\tau$ describes the inelastic scattering.
For elastic scattering, $\hbar /\tau$ can be both much smaller and much larger than temperature $T$.  
Moreover, some systems are not sensitive to the crossing the Planck dissipation criterion, $\hbar/\tau =T$.
Examples are provided by the $s$-wave superconductors with nonmagnetic impurities.
 According to the Anderson theorem,\cite{AG1959,Anderson1959}
in such superconductor, the transition temperature $T_c$ is the same as in the pure superconductor.
This is valid even in the highly super-Planckian regime:
 \begin{eqnarray}
E_F \gg \frac{\hbar}{\tau}  \gg T  \,,
 \label{superP}
\end{eqnarray}
where $E_F$ is Fermi energy.
Recently it was found that the Anderson theorem is also applicable to the polar phase of superfluid $^3$He -- the spin-triplet nodal-line superfluid with the $p$-wave pairing -- if impurities have the form of the columnar defects.\cite{Fomin2018,Fomin2020}
This extension of the Anderson theorem was confirmed in the experiments with the polar phase in nafen.\cite{Eltsov1908} 

However, the situation can be different for the other type of Planckian dissipation, in particular for the Planckian dissipation relevant for the vortex dynamics. In this case  the Planckian dissipation corresponds to the situation, when the inverse elastic scattering time $\hbar/\tau$ is compared with the distance $\Delta E$ between the Caroli-de Gennes-Matricon energy levels\cite{Caroli1964} in the vortex core: 
 \begin{eqnarray}
\frac{\hbar}{\tau} \sim \Delta E  \,.
 \label{PlanckDiss}
\end{eqnarray}
In this case the Planckian dissipation has the clear physical consequence: it determines the transition between the laminar and turbulence regime in the vortex dynamics.\cite{Finne2003,Finne2006}  The transition from super-Planckian to the sub-Planckian dissipation marks the onset of the vortex turbulence.

In the weak-coupling BCS regime $\Delta \ll E_F$,
where $\Delta$ is the superconducting gap, the vortex core contains the chiral branch of the discrete levels, which connects positive and negative energy states. These Caroli-de Gennes-Matricon energy levels are quantized with either $E_n=(n +1/2)\,\Delta E$ or 
$E_n= n\,\Delta E$   depending on the type of the Cooper pairing.\cite{Volovik1999} In the latter case the core contains the states with zero energy, $E_{n=0}=0$, which correspond to the Majorana fermions.\cite{Kopnin1991,Volovik1999,Ivanov2001} In the weak-coupling BCS regime, one has $\Delta E \sim \Delta^2/E_F \ll \Delta$, and the interlevel distance $\Delta E$ is called the minigap.

The scattering time $\tau$ related to the vortex may come either from the interaction of the core states with the bulk fermions, which play the role of the environment, or from the impurities, which also belong to the environment.  In both cases the scattering leads to the broadening of the core levels. If the broadening of the levels $\hbar/\tau$ becomes comparable with or larger than the minigap 
$\Delta E$, the energy levels in the vortex core overlap, and the chiral branch $E_n$ becomes continuous. This allows the spectral flow from the negative energy states to the positive energy  states, which leads to the spectral flow force acting on the vortex (the Kopnin force).\cite{Kopnin1995} This phenomenon represents the analog of the axial anomaly. The analogy becomes exact in case of continuous vortices (skyrmions) in $^3$He-A with Weyl fermionic quasiparticles. In this case, the spectral flow is governed by the Adler-Bell-Jackiw anomaly equation.\cite{Bevan1997} 
The experiment with skyrmions in the Weyl superfluid\cite{Bevan1997}  is probably the only reliable experiment in condensed matter, where the direct analog of the axial anomaly was experimentally observed and the measured prefactor in the Adler-Bell-Jackiw equation fully agreed with the number of the Weyl species in the chiral superfluid $^3$He-A (two Weyl species corresponding to two spin degrees of freedom).

It appears that the spectral flow is the important ingredient of the vortex dynamics, which was experimentally studied in detail both in the fully gapped $^3$He-B and in the chiral  $^3$He-A with Weyl fermions.\cite{Bevan1997,Bevan1997b} 
The effect of this peculiar dynamics on the vortex turbulence has been experimentally studied in $^3$He-B.\cite{Finne2003,Finne2006} 
Theoretically it follows from the coarse-grained hydrodynamic equation for the velocity of the superfluid component, which contains quantized vortices:
\begin{eqnarray}
\frac{\partial{\bf v}_s}{\partial t} +\nabla \mu ={\bf v}_s \times \boldsymbol\omega_s 
- \alpha' ({\bf v}_s -{\bf v}_n )\times \boldsymbol\omega_s
+ 
\nonumber
\\
+\alpha  \hat{\boldsymbol\omega}_s  \times [\boldsymbol\omega_s\times({\bf v}_s -{\bf v}_n )] \,\,\,,\,\,  \boldsymbol\omega_s=\nabla\times {\bf v}_s
\,,
\label{Core1}
\end{eqnarray}
where  ${\bf v}_s$ and ${\bf v}_n$ are the superfluid and normal velocities correspondingly, and $\hat{\boldsymbol\omega}_s$ is the unit vector of superfluid vorticity $\boldsymbol\omega_s$. 

The first term in the rhs of Eq.(\ref{Core1}) comes from the conventional Magnus force 
acting on vortices. The second term with the parameter $\alpha'$ is also the reactive (nondissipative) force, but it comes from the effect of the chiral anomaly. 
The third term with the parameter $\alpha$ is the mutual friction force between the superfluid and normal components, which is actually the friction force acting on quantized vortices, when they move with respect to the normal bath. 

 Vortex turbulence (or quantum turbulence) is the analog of the conventional turbulence in normal liquids, but the vorticity is provided by the distributed vortices (with quantized circulation). 
As in classical liquids the onset of quantum turbulence is governed by the corresponding Reynolds number, which is determined
by the ratio the reactive and dissipative terms.
However, as distinct from the turbulence in classical liquids the turbulence in the two-fluid hydrodynamics is described by 3 Reynolds numbers. One Reynolds number describes the dynamics of the normal component. In liquid $^3$He, due to the high viscosity of the normal component, its Reynolds number is small and the normal flow is laminar. It is typically at rest, i.e. ${\bf v}_n=0$.
Then the superfluid dynamics in Eq.(\ref{Core1}) contains two dimensionless parameters:
\begin{eqnarray}
1- \alpha' = \frac{\omega_0^2\tau^2}{1+\omega_0^2\tau^2} {\rm tanh} \frac{\Delta}{2T}\,,
\label{reactive}
\\
\alpha =  \frac{\omega_0\tau}{1+\omega_0^2\tau^2} {\rm tanh} \frac{\Delta}{2T}\,,
\label{dissipative}
\end{eqnarray}
where $\hbar\omega_0=\Delta E$ is the minigap.

 \begin{figure}[htt]
 \includegraphics[width=\columnwidth]{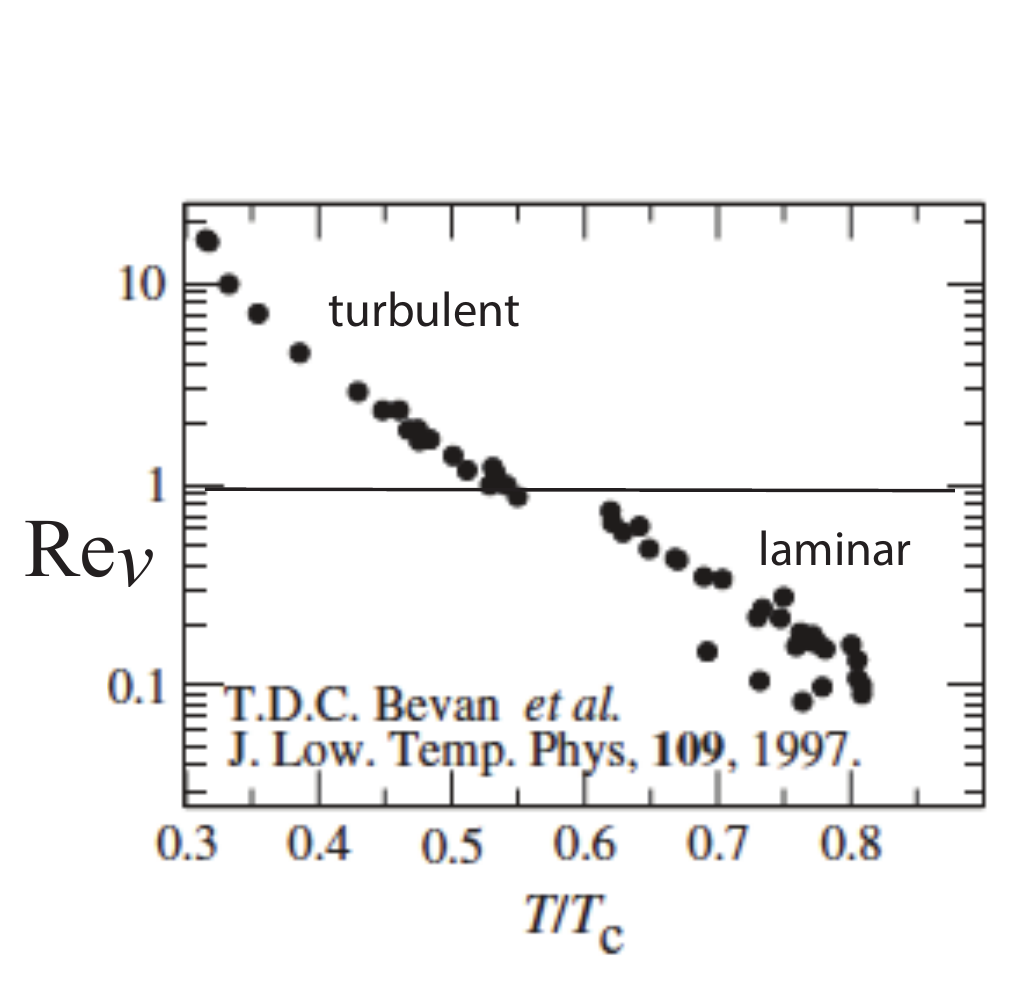}
 \caption{
Temperature dependence of the superfluid Reynolds number $Re_v$ with $1/Re_v= (\hbar/\tau)/\Delta E$, from experiments in $^3$He-B.\cite{Bevan1997b}
The critical Reynolds number $Re_v\sim 1$ marks the transition to superfluid turbulence
observed at $T/T_c \sim 0.6$, see Ref.\cite{Finne2003,Finne2006}. This critical superfluid Reynolds number corresponds to the Planck dissipation, $\hbar/\tau \sim \Delta E$.
}
 \label{TemperatureDependence}
\end{figure}

It is important that the  reactive and dissipative terms in Eq.(\ref{Core1})
have the same dependence on velocity and thus the superfluid Reynolds number describing the ratio of the two terms is velocity independent.
This superfluid Reynolds number (known as Kopnin number) has the direct relation to the  Planck dissipation in 
Eq.(\ref{PlanckDiss}): 
\begin{eqnarray}
Re_v =\frac{1-\alpha'}{\alpha} = \frac{\Delta E}{\hbar/\tau}
 \,.
\label{Rev}
\end{eqnarray}
This dimensionless Reynolds number is the function of temperature, and its experimental temperature dependence\cite{Bevan1997b} is shown in Fig.(\ref{TemperatureDependence}).

In the regime $Re_v <1$ the spectral flow along the core levels becomes allowed, and in the limit $Re_v\ll 1$ the Kopnin spectral force almost completely compensates the conventional Magnus force. In this regime, the dissipation wins over the combined reactive force, and this prevents the turbulence. 
In the other regime, when  $Re_v>1$, the energy levels become isolated from each other, and the spectral flow is suppressed. The Magnus force wins over dissipation giving rise to the quantum turbulence. 

The discussed Planckian dissipation at $Re_v\sim 1$ is similar to that in the mesoscopic systems, when the broadening  of the levels due to dissipation becomes comparable with the distance between the levels.\cite{Altshuler1986}

 \begin{figure}[htt]
 \includegraphics[width=\columnwidth]{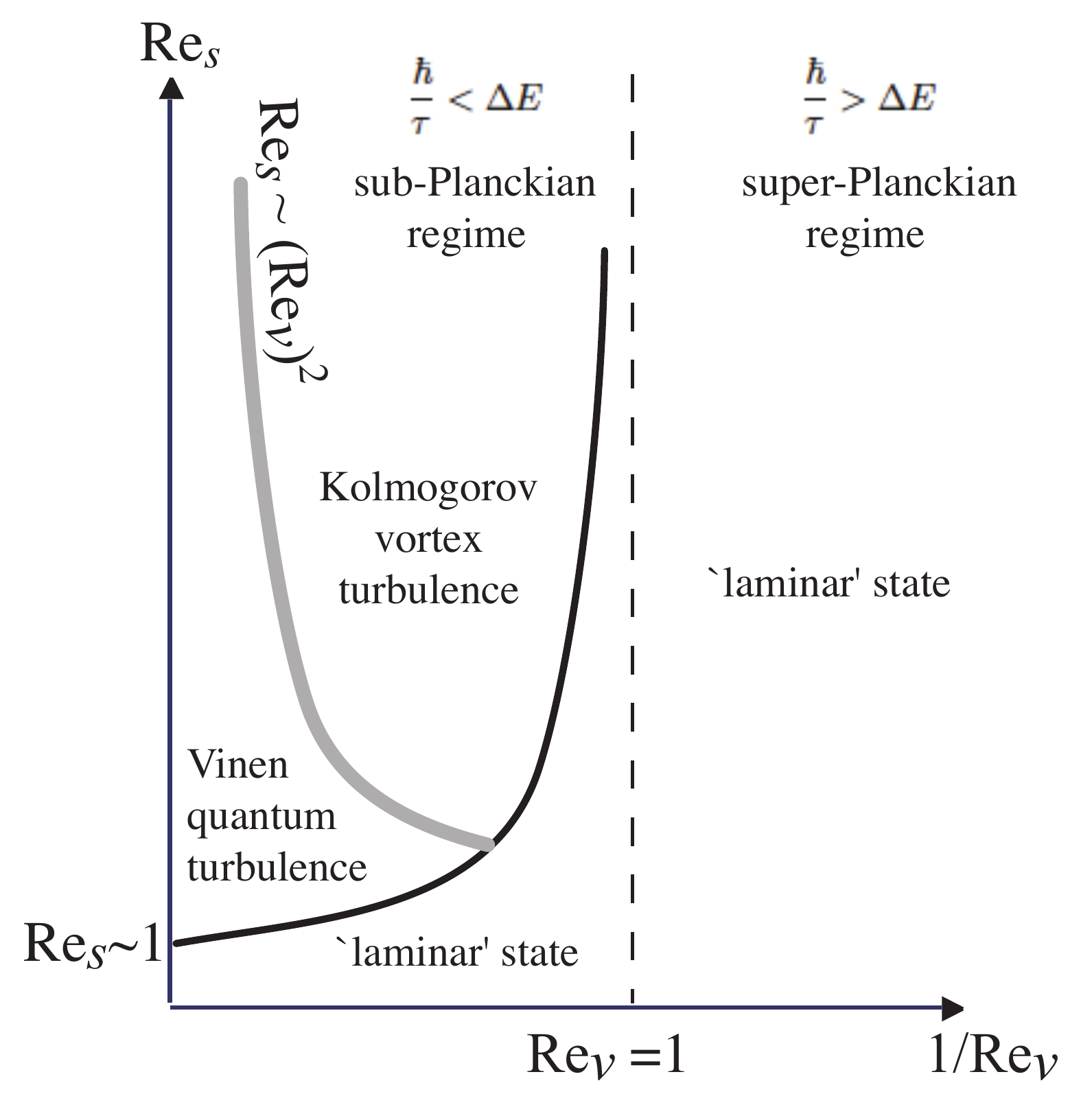}
 \caption{
 Phase diagram of the vortex turbulence from Ref.\cite{Volovik2003}
  in terms of two Planckian parameters, which serve as the quantum Reynolds numbers, $Re_v$ and $Re_s$, where $1/Re_v= (\hbar/\tau)/\Delta E$ and 
 $1/Re_s= \hbar/(mUR)$. Here $\Delta E$ is the distance between the levels in the vortex core; $m$ is the mass of $^3$He atom; $R$ is the size of the vessel; $U$ is the macroscopic velocity of the flow. The black solid line separates laminar and turbulent regimes.The vertical dashed line separates the sub-Palnckian dissipation with 
 $\hbar/\tau < \Delta E$ and  the super-Palnckian dissipation with $\hbar/\tau > \Delta E$, in which the effect of the axial anomaly takes place.
 For the sufficiently large $Re_s \gg 1$, the Planckian dissipation, $\hbar/\tau \sim \Delta E$, separates the laminar flow in the super-Planckian regime from the vortex turbulence in the sub-Palnckian regime.  The grey line marks the crossover between two types of vortex turbulence: the Vinen quantum regime with the single length scale, and the regime with the classical-like Kolmogorov cascade. }
 \label{TurbRegimes}
\end{figure}

The phase diagram of vortex turbulence is illustrated in Fig. \ref{TurbRegimes}. It is determined by two superfluid Reynolds numbers, $Re_v$ and $Re_s$, both can be called Planckian. The Reynolds number $Re_s$ is $Re_s= UR/(\hbar/m)$, where $U$ and $R$ are the parameters of the flow at large scale. This superfluid Reynolds number looks similar to the Reynolds number in classical liquids, $Re= UR/\nu$, but instead of the kinematic viscosity $\nu$ the quantum quantity $\hbar/m$ enters. This has some connection to the minimal viscosity.\cite{Policastro2001,Kovtun 2005,Zaanen2019}

The phase diagram in Fig. \ref{TurbRegimes} has three regions. At large enough temperature there is the region  of the laminar flow.  Experimentally the range of temperatures is $0.6 T_c < T < T_c$. At  $T\sim 0.6T_c$, the transition to the vortex turbulence takes place, which follows both from experiments and from the theoretical calculations of the transition temperature. Below this transition the vortex turbulence is described by the classical-like Kolmogorov cascade. The reason for that is that the vortex lines are locally polarized, forming the "classical" vortex tubes. The quantum effects related to the quantization of circulations are suppressed, and vortices can be described by the classical-like coarse-grained vorticity
 $\boldsymbol\omega_s$ in Eq.(\ref{Core1}). 
 
 This equation (\ref{Core1}) is not applicable at very low temperature, where the dynamics of the individual vortices becomes important. In this regime, the vortex turbulence is characterized by the single length scale -- the distance between  quantized vortices. This type of turbulence is known as Vinen turbulence.\cite{Vinen1961,Vinen2002} In Fermi superfluids, the crossover between Kolmogorov and Vinen turbulence takes place only at very low temperature, when $Re_s > Re_v^2$. On the contrary, in Bose superfluids the vortex turbulence is typically of the Vinen type, which is described in terms of the density of the vortex tangle $L \sim 1/a^2$, where $a$ is the distance between vortices.\cite{Schwarz1988,Aarts1994,Nemirovskii2008,Nemirovskii2020}

\section{Conclusion}

In conclusion, the notion of the Planckian dissipation can be extended to the system of the  Caroli-de Gennes-Matricon energy levels in the vortex core of superconductors and fermionic superfluids. In this approach the Planck dissipation takes place when the scattering time 
$\tau$ is comparable with the Heisenberg time $t_H=\hbar/\Delta E$, where $\Delta E$ is  the interlevel disctance (the minigap). 

The Heisenberg type Planck dissipation has two important physical consequences.  First, it determines the regime, when the effect of the axial anomaly becomes important. 
The anomalous spectral flow of levels along the chiral branch of the Caroli-de Gennes-Matricon states takes place in the super-Planckian region, i.e. when $\tau <\hbar/\Delta E$, and is absent  in the sub-Planckian region, $\tau>\hbar/\Delta E$. Second, the Planck dissipation separates the laminar flow of the superfluid liquid at $\tau<\hbar/\Delta E$ and the vortex turbulence flow at $\tau>\hbar/\Delta E$.

  {\bf Acknowledgements}.  This work has been supported by the European Research Council (ERC) under the European Union's Horizon 2020 research and innovation programme (Grant Agreement No. 694248).

\end{document}